\documentclass[superscriptaddress,twocolumn]{revtex4}
\pdfoutput=1
\usepackage{graphicx,color,url}
\definecolor{b}{rgb}{0,0,1.0}
\definecolor{r}{rgb}{1,0,0}
\definecolor{g}{rgb}{0,1,0}

\begin{document}

\newcommand{\SZFKI}{Research Institute for Solid State Physics and Optics,
 P.O. Box 49, H-1525 Budapest, Hungary}

\newcommand{\HASBUTE}{HAS-BUTE Condensed Matter Research Group,
Budapest University of Technology and Economics, H-1111 Budapest, Hungary}

\newcommand{\MDB}{Institute of Experimental Physics, Otto von Guericke University,
Universit\"atsplatz 2, D-39106 Magdeburg, Germany}

\newcommand{\MDBL}{Leibniz Institute for Neurobiology, D-39118 Magdeburg, Germany}

\title{Reflection and Exclusion of Shear Zones in Inhomogeneous Granular Materials}

\author{Tam\'as B\"orzs\"onyi}

\email{btamas@szfki.hu}
\affiliation{\SZFKI}
\author{Tam\'as Unger}
\affiliation{\HASBUTE}
\author{Bal\'azs Szab\'o}
\affiliation{\SZFKI}
\author{Sandra Wegner}
\affiliation{\MDB}
\author{Frank Angenstein}
\affiliation{\MDBL}
\author{Ralf Stannarius}
\affiliation{\MDB}

\begin{abstract}
Shear localization in granular materials is studied
experimentally and numerically. The system consists of two material
layers with different effective frictions. The presence of the
material interface leads to a special type of ``total internal
reflection'' of the shear zone. In a wide range of configurations
the reflection is characterized by a fixed angle which is analogous
to the critical angle of refraction in optics. The zone leaves and
reenters the high friction region at this critical angle and in
between it stays near the interface in the low friction region.
The formalism describing the geometry of the shear zones and that of
refracted and reflected light beams is very similar.  For the internal
visualization of shear localization two independent experimental
techniques were used (i) excavation and (ii) Magnetic Resonance
Imaging.

\vspace{0.5cm}
\hspace{-2.2cm}
\noindent Final version published in Soft Matter, DOI:10.1039/C1SM05762F (2011)   

\end{abstract}
\maketitle

\section{Introduction and model}

Shearing of complex materials e.g. non-newtonian fluids, colloids,
emulsions, foams or granular materials often leads to shear localization
\cite{dhle2003,chzu1992,bewe2007,sche2010,fehe2003,feme2004}. 
The region with the highest shear rate is called the shear zone (or shear band)
\cite{riwo2007,toun2007,feme2006,un2007,un2007NP,boun2009,knbe2009,lu2004,chle2006,dele2007,desa2006,safe2008,ja2008,mude2000,gdrmidi2004}.
Experiments and numerical simulations for granular materials
showed that the shear zone can exhibit nontrivial curved or refracted
shapes even for the simplest stationary case, where the
position and shape of the shear zone remains more or less the same during
the shear deformation \cite{sche2010,riwo2007,toun2007,un2007NP,feme2004,un2007,feme2006,boun2009,knbe2009}.
The case of inhomogeneous materials is especially important for understanding
various natural phenomena or industrial problems. In such systems, where the internal friction
of the material changes with location, the shear is preferentially localized
to regions with lower friction. A simple example is the case of a layered material
where different layers correspond to different friction $-$ a configuration which
is often seen in nature. The aim of the current paper is to characterize
the puzzling shapes of shear zones in such systems by experimental and numerical techniques.

The shape of the shear zone can be captured by a simple variational model
\cite{unto2004,un2007} where the basic idea is the following: the shear
zone is modeled as an infinitely thin sliding surface. For any potential
sliding surface that is consistent with the boundary conditions, one has
to determine its total shear resistance against the driving. The actual shear
zone will correspond to the sliding surface that has the lowest total
resistance. This is the weakest surface that yields first and separates the
system into two solid blocks that slide next to each other.
The protocol for the determination of the shear resistance depends on
the actual configuration.
On one hand, for Couette or cylindrical split-bottom shear cells the potential sliding
surfaces are loaded by the same mechanical {\em torque}. Therefore the weakest
surface has to be determined based on the ability to transmit the
torque. On the other hand, if straight shear cells are used as in the
present study, the potential sliding surfaces experience an equal shear
{\em force}. Therefore the shear zone will correspond to the potential sliding
surface that is the weakest regarding its ability to transmit the total shear
force between its two sides. In both cases -- described by minimum torque or
minimum force -- the shear deformation corresponds to the lowest
possible energy dissipation \cite{unto2004,un2007}.

\begin{figure*}[htb]
\includegraphics[width=18cm]{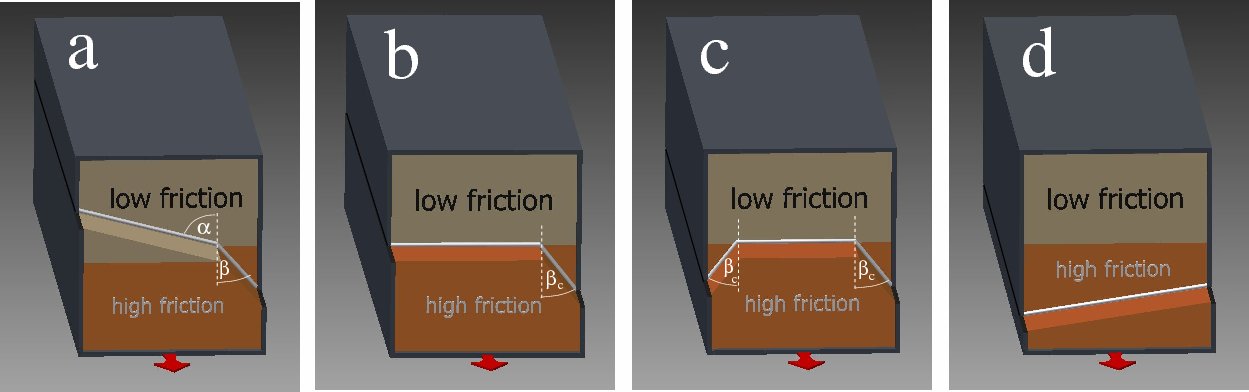}
\caption{
A gedankenexperiment: a material consisting of two layers is sheared
by moving the bottom part of the cell according to the red arrow.
The four panels illustrate how the boundary conditions affect the
position of the shear zone in this closed system. From (a) to (d) the left 
boundary of the zone (position of the slider edges) is moved downwards step 
by step, whereas its right boundary is kept fixed. The case (a) is analogous 
to the refraction of light beams. (b) Limit angle of refraction $\beta_c$.
Panel (c) shows the ``total reflection'' of the shear zone. It is quite
different from the total reflection of geometric optics as part of the zone
is excluded from the high friction material and the zone arrives at the
interface at the limit angle $\beta_c$. (d) Straight zone, not influenced by
the interface.
}
  \label{geometries}
\end{figure*}

In this paper we deal with straight shear cells where shearing is
provided by the relative motion of two long parallel sliders
(Figs.~\ref{geometries} and \ref{open-closed}). Then the shear zone
is created between two
material blocks, each stuck to one of the sliders. The location of the emerging
shear zone can be determined by the above consideration. As the
geometry is translation invariant in the direction of shear ($x$
direction) the shear zone can be represented by a single path in the cross
section of the cell ($yz$ plane). The total shear force that can be
transmitted by such a path is proportional to the integral

\begin{equation}
   \int \tau \ ds = \text{min.}  \, ,
\label{force}
\end{equation}

where $\tau$, the maximum shear stress that the material can
exert locally, is integrated along the length $s$ of the path. The shear
zone corresponds to the path for which the integral takes its minimum
value.  It is assumed that $\tau$ is given by the product of the
effective friction coefficient $\mu$ and the local
pressure\footnote{We neglect here, that the stress tensor is not
spherical.} $p$

\begin{equation}
   \tau = \mu p \, .
\label{shearstress}
\end{equation}

In an inhomogeneous system the effective friction $\mu$ can vary from
place to place, e.g. when layers of different materials are present.
When $p$ is constant, the condition (\ref{force}) reduces to $\int \mu
\ ds=\text{min.}$, which is formally identical to the well known
Fermat principle in optics. According to this, the light beam traveling
through an optically inhomogeneous material finds the optimal way,
where the length weighted by the refraction index is extremal. This
results in refraction of light beams at the boundaries of regions with
different optical indices. The analogy leads to the idea that
similar refraction effects are expected in these two distant fields of
physics. The case of zone refraction can then be described by a law
which is similar to the refraction law in optics

\begin{equation}
   \frac{\sin\alpha}{\sin\beta} = \frac{\mu_h}{\mu_l},  \nonumber
\end{equation}

where $\alpha$ and $\beta$ are the two angles of incidence (see Fig.~\ref{geometries}a),
while $\mu_l$ and $\mu_h$ stand for the effective friction in the two layers with low friction
and high friction, respectively. This law was recently tested numerically and experimentally
\cite{un2007,un2007NP,boun2009,knbe2009}.

 In the present work we do not deal with the case of refraction
  but focus on configurations in which the zone after visiting the
  interface eventually returns to the high friction part of the
  sample. The boundary conditions, where both ends are located in the
  high friction material, correspond to the total internal reflection
  in geometric optics. However, in this situation the behavior of the
  shear zone differs from the behavior of light beams
  (Fig.~\ref{geometries}c). Our version of Fermat's principle,
  i.e. the global minimum of $\int \mu \,\text{d}s$, suggests the
  following interesting scenario. Here the middle part of the shear
  zone is excluded from the high friction material and it is located
  as close as possible to the interface but in the low friction
  region. Thus this part of the zone enters with the coefficient $\mu_l$ into the
  calculation of the total transmitted shear force. The shear zone
  from both sides reaches the interface at the critical angle
  $\beta_c$ defined by $\sin\beta_c=\mu_l/\mu_h$. Interestingly this
  angle characterizes a wide range of configurations not only the
  limit case of refraction (Fig.~\ref{geometries}b).

It can be easily verified that the above exclusion effect and the
piecewise straight shape of the shear zone (Fig.~\ref{geometries}c)
indeed correspond to the least possible shear force within the
framework of the variational model of the infinitely narrow
zone. In the following we test this picture experimentally.

\section{Experimental methods}

\begin{figure}[ht]
\includegraphics[width=\columnwidth]{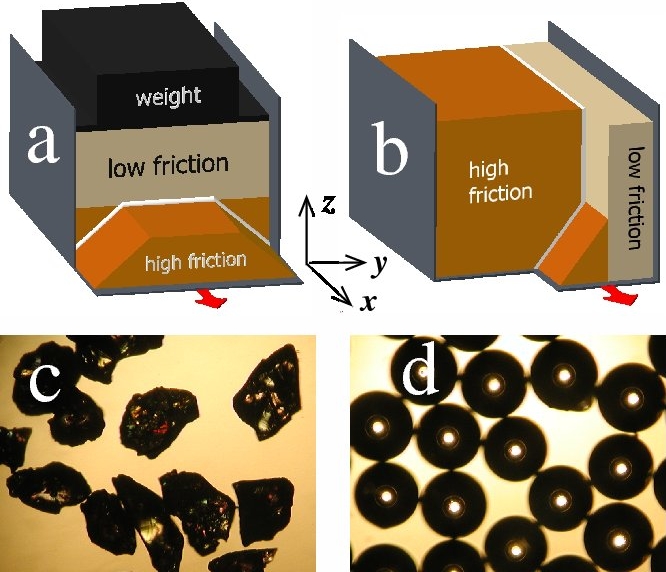}
\caption{(a)-(b) Schematic illustration of the two experimental configurations used.
    The layered granular material is sheared by moving (a) the lower boundary for the
    {\it "closed system"} or (b) one of the L shaped walls for the {\it "open system"}
    according to the red arrow.
    The location of the shear zone is indicated with a white line.
    The upper surface of the {\it "closed system"} is loaded with a weight 3 times
    heavier than the granular material below. (c)-(d) Microscopic
    images of the materials used in the experiments (c) corundum (grain size
    $d=0.33\pm0.02$\,mm) and (d) glass beads (grain size $d=0.48\pm 0.02$\,mm).
     }
  \label{open-closed}
\end{figure}

For the experimental realization of the above described systems we use
two materials with different frictional properties. One material is
corundum which consists of angular grains (see
Fig.~\ref{open-closed}c).  Therefore it has higher effective internal
friction than the other material consisting of glass beads
(Fig.~\ref{open-closed}d).  To characterize the difference in the
effective frictions, the angles of repose ($\theta_r$) were determined
by the method used in \cite{boha2008} and were
$\theta_r^{gla}=21.9^\circ$ for glass beads while $\theta_r^{cor} =
33.2^\circ$ for corundum. The latter is somewhat higher than the
typical value for sand ($\theta_r^{san} = 30.5^\circ$). The
effective frictions are simply estimated by
$\mu \approx \tan\theta_r$.  The ratio
$\tan\theta_r^{cor}/\tan\theta_r^{gla} = 1.63$ gives a reasonable
contrast for the effective friction.

Two experimental geometries were investigated. In the first one
$-$ which is referred to as {\it "closed system"} (see Fig.~\ref{open-closed}a) $-$
the zone was forced to start and end in the high friction part of the sample. Here
the granular materials were layered horizontally with the low friction
material placed on top of the high friction material and shearing was
obtained by a slow translation of the bottom plate according to the
red arrow. If the thickness of the lower (high friction) layer is not too large
the zone escapes it and then comes back to it, as it is indicated by the white line.
In order to reduce the relative pressure difference
between the top and bottom of the system the upper surface was loaded
with a weight which was about 3 times heavier than the granular
material below.  The length of the cell was $70$ cm and it had an internal
cross section of $4.5$ cm $\times$ $4$ cm.

The second experimental apparatus referred to as {\it "open system"}
(see Fig.~\ref{open-closed}b) included two 70 cm long L shaped sliders
(cell wall), one of which was slowly translated in the experiments.
When the two granular layers (with low and high friction) are arranged
vertically (as seen on the the sketch), the shear zone takes a form as
it is indicated by the white line. It is forced to start in the high
friction part of the sample but is free to select the position of the
other end at the top surface. This cell had an internal cross section
of $4$ cm $\times$ $3.5$ cm.  The total shear displacement
achieved in each experiment was between $5$ and $6$ cm for both cases
(Figs.~\ref{open-closed}a and \ref{open-closed}b).

\begin{figure}[ht]
\includegraphics[width=\columnwidth]{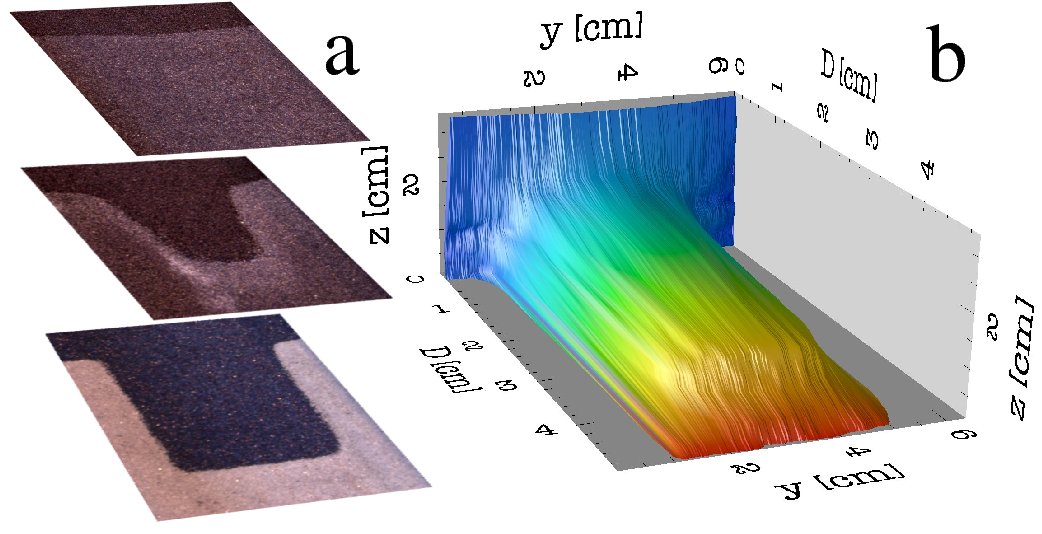}
\caption{
Demonstration of the excavation method for the case of the {\it closed system}.
(a) Sample images taken during excavation. The displacement of the grains inside the
sample is visualized by carefully removing the material layer by layer.
(b) Displacement $D(y,z)$. The position of the interface between the two materials
was at height $1.0$ cm.
}
  \label{demo-closed}
\end{figure}

\begin{figure}[ht]
\includegraphics[width=\columnwidth]{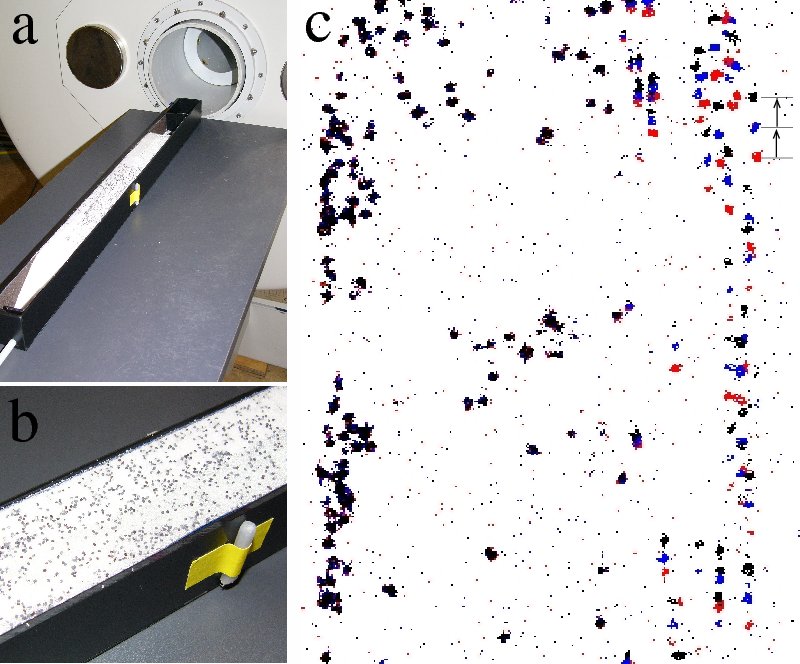}
\caption{
  Demonstration of the
  measurements using Magnetic Resonance Imaging for the case of the
  {\it open system}.  (a)-(b) The experimental cell filled with the
  material doped with poppy seeds.  (c) Overlapped horizontal sample
  images taken at $z=3.2$ cm. Poppy seeds appear as distinct spots
  on the binarized images with different colors for the 3 subsequent
  images.  As it is expected only the right hand
  side of the material is moving (see Fig.~\ref{open-closed}b), and the
  shear zone is shifted to the right with respect to the cell middle.}
  \label{demo-open}
\end{figure}

Two methods were used for the measurement of the deformation: (i) using colored
samples and excavating the material layer by layer (see Fig.~\ref{demo-closed})
the displacement was detected optically by digital imaging and (ii) the whole cell was
placed into a Magnetic Resonance Imaging (MRI) apparatus and the displacement
was detected by tracer particles (see Fig.~\ref{demo-open}).
In the following we demonstrate both methods.

In the case of optical detection we used two samples with different colors each for
both materials. In the upper layer, brown and gray correspond to glass beads
(grain sizes $d=0.56\pm0.02$\,mm and $d=0.48\pm 0.02$\,mm), while in the
bottom layer, dark blue and white correspond to corundum
(grain sizes $d=0.33\pm0.02$\,mm and $d=0.23\pm 0.02$\,mm).
After each experiment the material was carefully removed layer by layer
\cite{feme2006,boun2009} and the displacement profile $D(y)$ was determined for
each horizontal layer as it is illustrated by sample images in Fig.~\ref{demo-closed}a.
We used a commercial vacuum cleaner with additional extension tubes to provide a
slow flow rate. The procedure takes several hours for one experiment.
Typically 20-25 layers were recorded with the smallest interslice distance of $1$ mm.
The $D(y,z)$ displacement profile shown in Fig.~\ref{demo-closed}b resulted after combining
the information obtained for all slices.

Visualization of the internal deformation of the material by Magnetic Resonance
Imaging was realized using a Bruker BioSpec 47/20 MRI scanner
operating at 200 MHz proton resonance frequency (4.7 T) at the Leibniz
Institute for Neurobiology, Magdeburg. The cross section of our
experimental apparatus was optimized for the geometry of this device
with $7$ cm internal coil diameter.  Since the granular samples used
(corundum and glass beads) do not give an MRI signal, tracer particles
were needed. The tracer particles should be large enough allowing the
detection of single particles, but also not too large for the best
spatial resolution. The best signal was obtained using poppy seeds
with the size of $d=0.75\pm 0.04$\,mm). Similarly to the case of
excavation horizontal slices were obtained with the interslice
distance of $0.8$ mm and the in plane resolution of $0.156$
mm/pixel. The slider was displaced by $2.5$ mm between subsequent MRI
scans, Fig.~\ref{demo-open}c shows 3 subsequent images overlapped
taken at $z=3.2$ cm. During one experiment 20-30 displacement steps
were recorded yielding
a total displacement of $5-7.5$ cm.  The total measurement took about
4 hours. We note, that a potential segregation of the tracers during
shear if present, would be immediately evident in the MRI images. Such
segregation played no role in the present experiments.

\section{Simulation method}

The computer simulations presented here are based on the variational
narrow band model that we summarized in the introduction. However, the
simulation tries to overcome one serious drawback of the narrow band
model, namely, that it regards the shear zone infinitely thin. In
reality the zone has a non-negligible finite width and the system has a
smooth shear profile. In the simulation, wide shear zones are achieved by
introducing fluctuations due to disorder of the granular material as
follows. For a given state of the system, i.e. for one random realization
of the material, we identify the weakest sliding surface according to the
variational narrow band model (details are given below). This
procedure can be repeated many times for various random realizations
which results in a large number of narrow shear bands. Taking an ensemble
average over the random realizations provide us with a smooth shear
profile and a wide shear zone. A justification of this approach can
be seen in the three-dimensional structure of the actual shear zone. It adopts
an averaged profile in the $yz$ cell cross section by averaging over all slices
along the $x$ direction.

Such a protocol has been first used in computer simulations performed
by T\"or\"ok et. al. \cite{toun2007}, however, there are two major
differences between those simulations and the present method.
First, here we use independent random realizations as
opposed to the self-organized random potential used earlier
\cite{toun2007}. Second, the square lattice that was used in the
previous studies \cite{toun2007} inevitably introduces preferred
orientations for the shear zone. It is important that we avoid such a
bias. Therefore here an isotropic random mesh is used instead
of a regular lattice.

The two-dimensional random mesh is generated in the $yz$ cross
\begin{figure}[htb]
\includegraphics[width=\columnwidth]{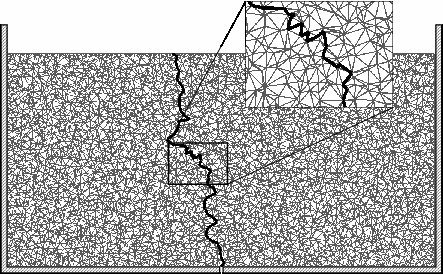}
\caption{
The random mesh used by the MC simulation and one shear
band. This situation corresponds to the straight split bottom cell filled
with one material. The band starts at the split line of the bottom and ends
at the free top surface of the system. The shear band represents the weakest sliding surface
for the given random realization of the material.}
  \label{mesh}
\end{figure}
section of the system perpendicular to the shear direction. First, we
start with the grid points of a square lattice (without bonds) of
lattice constant $a$. Second, the position of each point is
regenerated in a square of size $2a$ times $2a$ which is centered on
the original position of the point. The new random position is then chosen
uniformly inside the square. In the third step the points are
connected with bonds using Delaunay triangulation
(Fig.~\ref{mesh}). The resulting mesh is macroscopically isotropic to a good
approximation. For example testing the bond orientations shows that all
directions are equally probable and we found no sign of the
orientation of the original square lattice.

The triangles of the mesh correspond to microscopic blocks of the
material while bonds represent interfaces between the blocks where
sliding can take place.  The maximum shear force\footnote{In fact, the
quantity $S$ calculated here has the meaning of shear force per unit
length in $x$-direction.} $S$ that a given bond can exert without sliding
is given by $l \mu p R$, where $l$ is the length of the bond, $\mu$
is the local coefficient of the effective friction, $p$ is the local
pressure and $R$ is a random number chosen uniformly between 0 and 1.
Each bond has its own random number $R$ that is regenerated for each
random realization of the material. It is assumed that the local
pressure $p$ is hydrostatic, i.e. $p$ is proportional to the weight
of the material above.

Any potential sliding surface which is consistent with the boundary
conditions can now be represented by a continuous chain of
bonds. For example the two ends of the chain are fixed for the closed system
according to Fig.~\ref{open-closed}a, while for the open system
(Fig.~\ref{open-closed}b) only the lower end is fixed at the split
line of the boundary, the upper end is free, i.e. it can be anywhere
at the top surface of the material. For such a chain $\Gamma$ the
total shear force that it can resist is given by the sum of bond
forces $S_i$ for all the bonds $i$ contained in $\Gamma$.
The actual shear band $\Gamma$, i.e. the weakest sliding surface,
is determined by the condition
\begin{equation}
   \sum_{i \in \Gamma} l_i \mu_i p_i R_i = \text{min.}   \,
\label{MCforce}
\end{equation}
The shape of the shear band depends on the actual random
realization. Such a shear band represents a displacement jump and
therefore a noncontinuous displacement field. It divides the
material into two regions: The region on one side of the shear band
has a unit shear displacement in $x$-direction while the region on the
other side has zero displacement.

In our Monte Carlo simulation we generate shear bands for many random
realizations (typically a couple of thousand bands are collected).
The displacement field that is obtained by averaging over random
realizations is continuous and can be directly compared to the
experimental displacement profiles.

The local friction coefficients $\mu_i$ are set according to the
experimentally measured values. There is only one free parameter left
in the simulation, namely, the resolution of the mesh based on the
lattice constant $a$. The parameter $a$ has to be calibrated once for
each type of material. This has been done with the help of the
straight split bottom cell using only a single material and by
matching the width of the shear zone between simulation and
experiment. This provided us with calibration ratios between grain
size $d$ and lattice constant $a$. We obtained $a/d = 0.42$ for glass
beads, and $a/d=0.11$ for corundum.
This is in accordance with other measurements on homogeneous materials
\cite{feme2004}
showing that the thickness of the zone is larger for spherical beads than
for irregular grains.
This also means that in the
simulations of the inhomogeneous systems presented in this paper $a$
is not constant in space. The applied resolution of the underlying
mesh is different for the different materials. After this calibration
is done there is no fit parameter left in the simulation method.

\section{Results}

\begin{figure}[ht]
\includegraphics[width=\columnwidth]{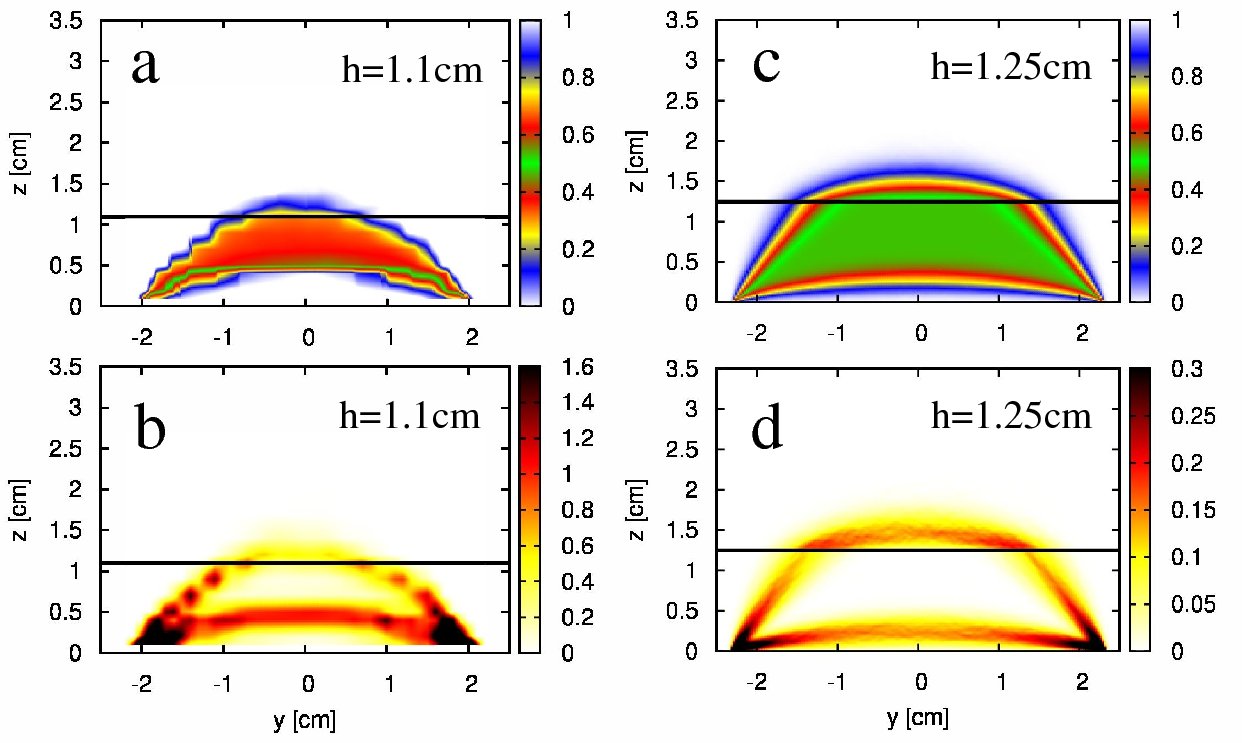}
\caption{Splitting of the zone.  (a) Displacement $D(y,z)$ and (b)
  strain in the material in the {\it closed system} for $h=1.1$ cm of the
  position of the interface, measured by excavation.
  (c)-(d) Results of numerical simulations obtained by the fluctuating
  narrow band model. In the simulation the splitting of the shear
    zone occurs at a somewhat higher position of the interface ($h=1.25$ cm).
    }
\label{split}
\end{figure}

We first present the results obtained for the {\it closed system} using excavation.
\begin{figure}[ht]
\includegraphics[width=\columnwidth]{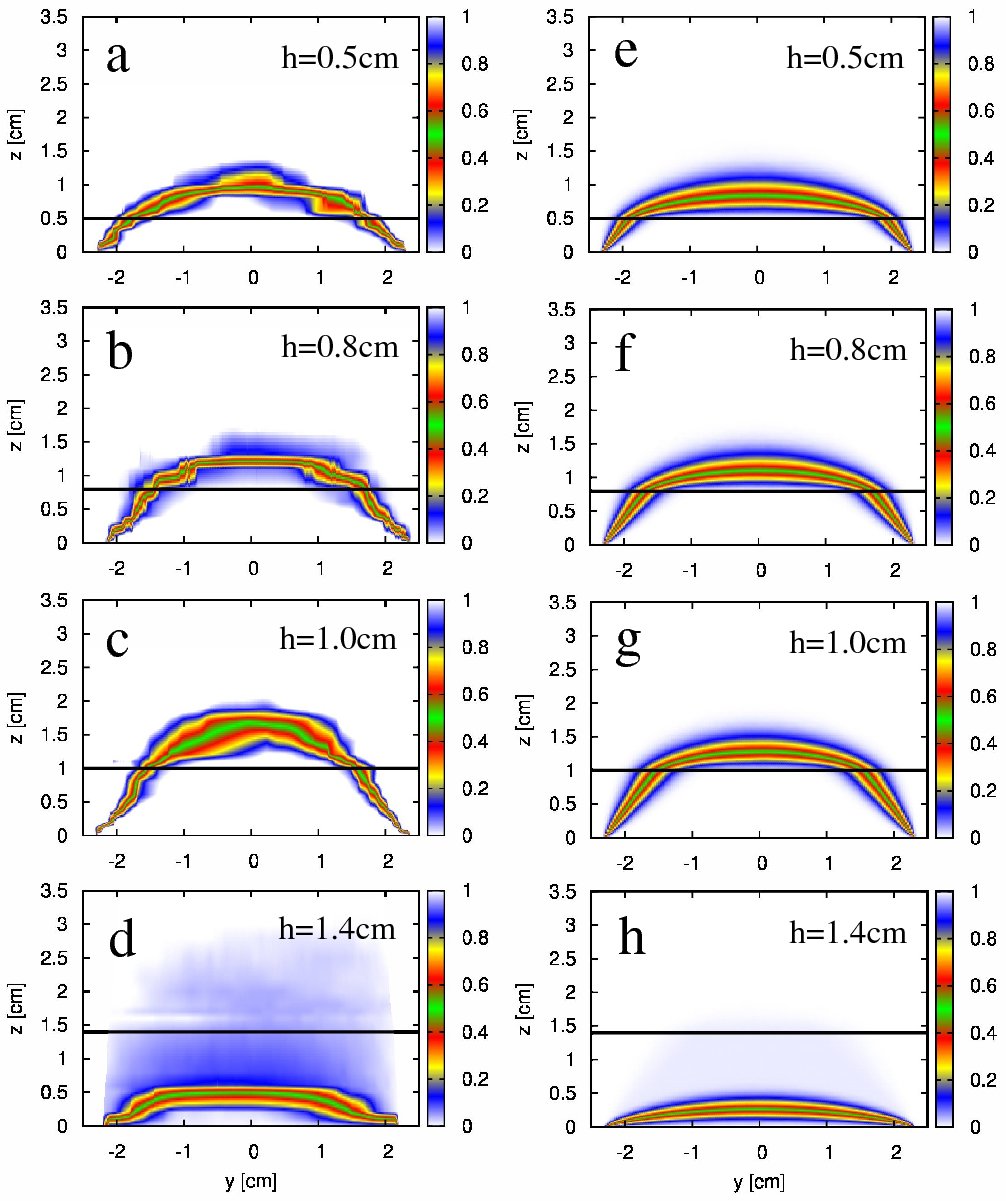}
\caption{(a)-(d) Displacement $D(y,z)$ of the material in the {\it closed system} for
four values of the position of the interface $h=0.5$ cm, $h=0.8$ cm, $h=1.0$ cm and
$h=1.4$ cm measured by excavation.
(e)-(h) Results of numerical simulations obtained by the fluctuating narrow band
model for the same configurations.}
\label{closedres}
\end{figure}
The final displacement $D(y,z)$ of the material is presented in
Figs.~\ref{closedres}a-d for four measurements in
such a way, that regions with highest shear deformation are visualized.
As it is seen both $-$ the experimental data and the
numerical calculations (Figs.~\ref{closedres}e-h) $-$ confirm, that for a thin enough corundum layer
the shear zone escapes the corundum and stays excluded from the corundum
just above the interface and then returns to the corundum to reach the
other corner of the setup. For a critical thickness $h_c$ of the corundum layer the
expression (\ref{force}) has two minima of equal shear force, meaning that there are two optimal
configurations: (i) the zone stays in the lower layer or (ii) the zone
escapes the lower layer as we have seen for thin layers. These two
configurations correspond to the same total force (Eq.~(\ref{force})).
For homogeneous pressure this yields 

\begin{equation}
h_c=\frac{W}{2}\sqrt{\frac{\mu^h-\mu^l}{\mu^h+\mu^l}},
\nonumber
\end{equation}

where $W$ stands for the width of the cell. For the given cell thickness the estimation
for $h_c$ would be $1.1$ cm.
A similar, but somewhat more complicated calculation that includes hydrostatic
pressure and the density difference between corundum and glass beads yields $h_c\approx 1.3$ cm.
For thick corundum layers the zone stays in the corundum, for which
an example is shown in Fig.~\ref{closedres}d at $h=1.4$ cm.
The observations show very nice quantitative agreement with the above calculations,
since we find that when $h$ is near the calculated critical thickness,
the zone splits up into two branches as it is seen in Fig.~\ref{split}.
This happens at around $h=1.1$ cm in the experiments and $h=1.25$ cm in the simulations.
The zone is remarkably bent in all
cases, which is apparently a direct consequence of the vertical pressure gradient
inside the granular material. Namely, frictional forces are smaller at
higher levels. Thus Eq.~(\ref{force}) results in a curved trajectory even
in a homogeneous layer.

\begin{figure}[ht]
\begin{center}
\includegraphics[width=6cm]{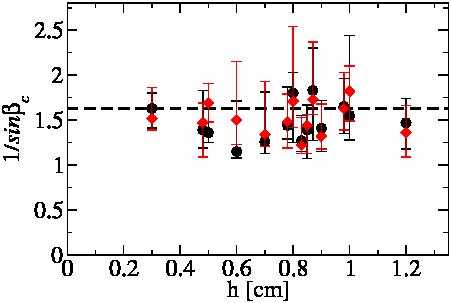}
\end{center}
\caption{The $sine$ of the angle of incidence $\beta_c$
as a function of the vertical position $h$ of the interface between
the two layers. Each measurement provides two data points, which are
marked with different symbols for clarity. The horizontal dashed line denotes
the ratio of the effective friction of corundum and glass beads estimated by angle
of repose measurements.}
\label{angle}
\end{figure}

Numerous experiments have been carried out using the {\it closed system}.
In Fig.~\ref{angle} we plot the $sine$ of the angle of incidence
as a function of the vertical position $h$ of the interface between
the two layers. As it is seen, the measured data match the
expected value $1/\sin\beta_c=\mu_h/\mu_l=1.63$ within experimental error.

\begin{figure*}[ht]
\begin{center}
\includegraphics[width=17cm]{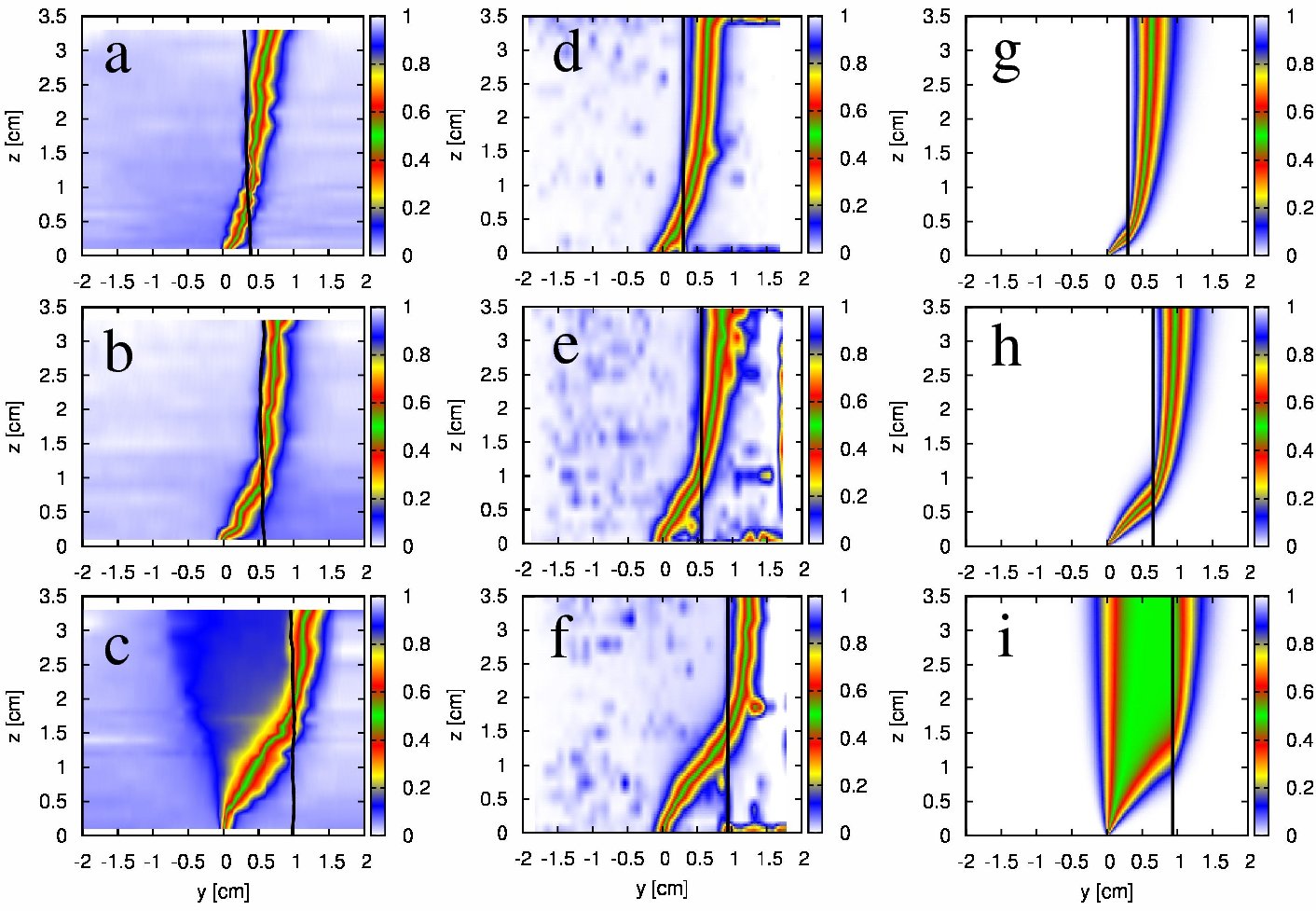}
\end{center}
\caption{Displacement $D(y,z)$ of the material in the {\it open system} for three values
of the position of the interface
$y=0.3$ cm, $y=0.66$ cm and $y=0.94$ cm measured by (a)-(c) excavation and (d)-(f)
Magnetic Resonance Imaging.
(g)-(i) Results of numerical simulations obtained by the fluctuating narrow band
model for the same configurations.}
\label{openres}
\end{figure*}

Experiments for the {\it open system} were carried out using both detection
methods: excavation and MRI.  Several configurations were
prepared with different position of the vertically aligned
interface. The results of three measurements are shown in
Fig.~\ref{openres} for $y=0.3$ cm, $y=0.66$ cm and $y=0.94$ cm. As it
is seen the zone starts from the middle of the cell and  moves up toward
the right hand side of the sample. Then it leaves the high friction part
and continues toward the top surface in the low friction part of the
sample.
We can calculate the critical distance $y_c$ at which zone splitting is expected
in a similar way as we did for the
closed system. Here we also take into account that pressure is zero at the top surface and
it increases linearly with depth.
This yields the critical distance

\begin{equation}
y_c = H \left(\sqrt{\left(\frac{\mu^h}{\mu^l}\right)^2-1}
- \sqrt{\left(\frac{\mu^h}{\mu^l}\right)^2-\frac{\mu^h}{\mu^l}}\right),
\nonumber
\end{equation}

which gives $y_c=0.93$ for the filling height of $H=3.4$ cm. The simulation
nicely reproduces the splitting of the zone for this configuration (see
Fig.~\ref{openres}i) and it is partly visible on the experimental data
(Fig.~\ref{openres}c).  In case the interface is further away from the
center ($y \gg y_c$) we obtain a vertical shear zone in the middle of
the cell (not shown here). Then the system behaves as if only the high
friction material was present and the interface has no effect on the
properties of the shear zone.

Finally, we discuss one difference in the MRI and excavation
techniques. While the MRI analysis provides the differential
displacement in individual steps, the excavation yields only the
integral displacement at the end of the experiment. If the zone
position and shape is stationary during the whole experiment, these
two informations are equivalent.  However, if an initial transient
corresponding to the shear zone formation is present, it affects only
the excavation data. As the comparison of Figs.~\ref{openres}a-c with
Figs.~\ref{openres}d-f shows, this influence plays no role for the
present experiments.  Nevertheless the quantitative characterization
of the initial transient behavior, which is in principle possible with
the MRI technique, will be an interesting aspect of future research.

\section{Summary}

Shear localization in inhomogeneous granular materials has been
studied experimentally and numerically. The shear zone preferentially
develops in regions with lower friction.  In systems consisting of
layers with different effective frictions the zone often changes
direction at the layer boundary $-$ a phenomenon analogous to light
refraction in geometrical optics. The formalism describing the
geometry of the shear zones and that of refracted and reflected light
beams is very similar.

Here we have shown that total internal reflection exists also for
shear zones. However, unlike in optics the zone reflection occurs
always at the critical angle of refraction $\beta_c$. In case of
shear zones this angle is defined by the ratio of the effective
frictions of the two material layers,
$\sin\beta_c=\mu_l/\mu_h$. This special reflection also involves
that a part of the shear zone is trapped at the interface of the
layers. We expect that such effects are present in naturally
layered systems consisting of not only two but multiple layers,
and could be best observed if the direction of shear velocity lies
parallel while its gradient is perpendicular to the layers.
Our measurements also demonstrate, that Magnetic Resonance
Imaging is a very powerful tool to accurately monitor the internal 
reorganization of granular materials in quasistatic flows.

\section*{Acknowledgments}
The authors are thankful for discussions with J\'anos Kert\'esz and for technical help from
Tilo Finger and Andr\'e Walther.
T.B. and T.U. acknowledge support from the Hungarian Scientific Research Fund
(Contract Nos.\ OTKA F060157 and PD073172).

\footnotesize{

}


\begin{thebibliography}{99}

\bibitem{dhle2003} J.K.G. Dhont, M.P. Lettinga, Z. Dogic, T.A.J. Lenstra and H. Wang,
Faraday Discuss. {\bf 123}, 157 (2003).

\bibitem{chzu1992} L.B. Chen, C.F. Zukoski, B.J. Ackerson, H.J.M. Hanley, G.C. Straty,
J. Barker and C.J. Glinka,
Phys. Rev. Lett. {\bf 69}, 688 (1992).

\bibitem{bewe2007} R. Besseling, E.R. Weeks, A.B. Schofield and W.C.K. Poon,
Phys. Rev. Lett. {\bf 99}, 028301 (2007).

\bibitem{fehe2003}  D. Fenistein, and M. van Hecke,
Nature {\bf 425}, 256 (2003).

\bibitem{sche2010} P. Schall and M. van Hecke, Annu. Rev. Fluid Mech. {\bf 42}, 67 (2010).

\bibitem{feme2004} D. Fenistein, J.W. van de Meent and M. van Hecke,
Phys. Rev. Lett. {\bf 92}, 094301 (2004).

\bibitem{riwo2007} A. Ries, D.E. Wolf and T. Unger,
Phys. Rev. E {\bf 76} 051301 (2007).

\bibitem{toun2007} J. T\"or\"ok, T. Unger, J. Kert\'esz, and D.E. Wolf,
Phys. Rev. E {\bf 75}, 011305 (2007).

\bibitem{feme2006} D. Fenistein, J-W. van de Meent and M. van Hecke,
Phys. Rev. Lett. {\bf 96}, 118001 (2006).

\bibitem{un2007} T. Unger,
Phys. Rev. Lett. {\bf 98}, 018301 (2007).

\bibitem{un2007NP}
Nature Physics, Shear light refraction, Research Highlights, {\bf 3}, 76 (2007).

\bibitem{boun2009} T. B\"orzs\"onyi, T. Unger and B. Szab\'o,
Phys. Rev. E {\bf 80}, 060302(R) (2009).

\bibitem{knbe2009} H.A. Knudsen and J. Bergli, Phys. Rev. Lett. {\bf 103}, 108301 (2009).

\bibitem{lu2004} S. Luding, Molecular Dynamics Simulations of Granular
  Materials, in The Physics of Granular Media, ed. Hinrichsen H. and Wolf
  D.E., p. 299 (Wiley, 2004).

\bibitem{chle2006} X. Cheng, J.B. Lechman, A. Fernandez-Barbero, G.S. Grest, H.M. Jaeger,
G.S.  Karczmar, M.E. Mobius and S.R. Nagel,
Phys. Rev. Lett. {\bf 96}, 038001 (2006).

\bibitem{dele2007} M. Depken, J.B. Lechman, M. van Hecke and W. van Saarloos,
Europhys. Lett. {\bf 78}, 58001 (2007).

\bibitem{desa2006}
M. Depken, W. van Saarloos and M. van Hecke,
Phys. Rev. E {\bf 73} 031302 (2006).

\bibitem{safe2008} K. Sakaie, D. Fenistein, T.J. Carroll, M. van Hecke and P. Umbanhowar,
Europhys. Lett. {\bf 84}, 38001 (2008).

\bibitem{ja2008} E.A. Jagla, Phys. Rev. E {\bf 78}, 026105 (2008).

\bibitem{mude2000}
D.M. Mueth, G.F. Debregeas, G.S. Karczmar, P.J. Eng, S.R. Nagel and H.M. Jaeger,
Nature {\bf 406}, 385 (2000).

\bibitem{gdrmidi2004} GDR MiDi, Eur. Phys. J. E {\bf 14}, 341-365 (2004).

\bibitem{unto2004} T. Unger, J. T\"or\"ok, J. Kert\'esz and D.E. Wolf,
Phys. Rev. Lett. {\bf 92}, 214301 (2004).

\bibitem{boha2008} T. B\"orzs\"onyi, T.C. Halsey, R.E. Ecke,
Phys. Rev. E {\bf 78}, 011306 (2008).

\end{thebibliography}
\end{document}